\documentclass[pra,showpacs,twoside,twocolumn]{revtex4}
\usepackage{amsmath,amsfonts,graphicx}
\usepackage[amssymb]{SIunits}
\usepackage[active]{srcltx}
\usepackage{color}

\newcommand{\ket}[1]{\left|#1\right\rangle}

\newcommand{\bra}[1]{\left\langle#1\right|}
\newcommand{\bracket}[2]{\left\langle#1|#2\right\rangle}

\DeclareMathOperator{\tr}{Tr} 
\DeclareMathOperator{\id}{\openone}
\SetSymbolFont{symbols}{normal}{OMS}{cmsy}{m}{n}

\begin{document}
\bibliographystyle{apsrev}
\title{Einstein-Podolsky-Rosen correlations in a hybrid system}

\author{Pawe{\l}{} Caban}
\email{P.Caban@merlin.phys.uni.lodz.pl}
\author{Jakub Rembieli{\'n}ski}
\email{J.Rembielinski@merlin.phys.uni.lodz.pl}
\author{P. Witas}
\email{pjwitas@gmail.com}
\author{M. W{\l}odarczyk}
\email{marta.wlodarczyk@gmail.com}

\affiliation{Department of Theoretical Physics, University of Lodz\\
Pomorska 149/153, 90-236 {\L}{\'o}d{\'z}, Poland}
\date{\today}
\begin{abstract}
  We calculate the relativistic correlation function for a hybrid
  system of a photon and a Dirac-particle.  Such a system can be
  produced in decay of another spin-$1/2$ fermion. We show, that the
  relativistic correlation function, 
  which depends on particle momenta, may have local extrema for
  fermion velocity of order $0.5c$. This influences the degree of
  violation of CHSH inequality.
\end{abstract}

 \maketitle

\section{Introduction}

In this paper we derive and study the Einstein--Podolsky--Rosen (EPR)
correlations in a hybrid system of a massive spin-1/2 fermion and the
photon in the relativistic regime.  Investigation of the relativistic
EPR correlation functions is important in view of certain
tension between the spirit of special relativity and non-locality of
quantum 
mechanics. This subject has been dealt with by numerous authors in
recent years \cite{Czachor1997_1, RS2002, GA2002, GBA2003, TU2003_1,
  TU2003_2, ALMH2003, Terno2003, CW2003, LD2003, CR2003_Wigner,
  LY2004, SL2004, PT2004_1, YWYNMX2004, Czachor2005, CR2005, PST2005,
  KS2005, CR2006, Caban2007_photons, Caban_2008_bosons_helicity,
  ja2008, ja2009, smolinski, CDKO_2010_fermion_helicity, FBHH2010,
  czachor2010, CRW_2010_nonscalar}.
These studies have shown that the EPR correlations of
relativistic massive particles differ from those of non-relativistic
ones, namely the correlations depend on particle momenta and may show
non-monotonic behavior. Moreover the latter are similar to the
correlations involving only photons. Thus it is strongly desirable to
measure the EPR correlations using relativistic massive particles
\cite{Caban2007_photons}.

To our knowledge, three experiments with the use of relativistic
massive fermions have been performed so far.  All the experiments
measured proton-proton correlations. The first one, known as
Lamehi--Rachti--Mittig experiment (LRM) \cite{LRM}, was performed in
CEN-Saclay about thirty years ago. The LRM group tested Bell
inequalities with the use of low-energy proton beams ($13,5$ MeV, that
is $v\sim0,17 c$).  The second experiment {was conducted} in
Kernfysisch Versneller Instituut by S. Hamieh group \cite{KVI}.  In
this experiment, spin correlations of proton pairs in $^1S_0$ state,
emerging from $^{12}\mathrm{C}(d,^2\mathrm{He})^{12}\mathrm{B}$
nuclear charge-exchange reaction, were examined. Energy of protons was
about $86$ MeV ($v\sim0.4 c$). The third experiment was performed in
RIKEN Accelerator Research Facility by H. Sakai group \cite{RIKEN}.
In the RIKEN experiment, proton pairs produced in
$^1\mathrm{H}(d,^2\mathrm{He})n$ reaction reached energy of $135$ MeV
($v\sim 0.5 c$). Results of these experiments are in agreement with
the predictions of non-relativistic quantum mechanics. This is not
surprising as the relativistic effects appear at much higher energies.
For proton they are of the order of $800$ MeV ($v\sim 0.85 c$)
\cite{ja2009, smolinski}. However, it is an unprecedented difficulty to
produce a singlet state and measure spin correlations of pairs of
energetic protons. Thus it is fully justified to search for other
processes that might allow a conclusive measurement. One of them are
correlations in a hybrid system, that is in a system consisting of
massive particles and photons.

In this paper we present a new concept to measure correlations in a
simple hybrid system consisting of a massive spin-$1/2$ fermion and a
photon. We show that for velocities $v\gtrsim 0.5c$ the fermion-photon
correlation function depends non-monotonically on particle
momenta. We also find that the violation of the
Clauser-Horne-Shimony-Holt (CHSH) inequalities depends on particle
momenta in non-monotonic way.

The paper is organized as follows. In Sec.~\ref{s:hybrid states} we
construct the hybrid states. In Sec.~\ref{s:observables} we define
observables of polarization (for photon) and relativistic spin (for
fermion). Then we calculate the correlation function
(Sec.~\ref{s:correlation function}) and analyze the CHSH inequalities
in some special cases (Sec.~\ref{s:bell}). Conclusions are given in
the Sec.~\ref{s:conclusions}.

\section{Construction of the state
 of the  hybrid system} 
\label{s:hybrid states}

We shall consider the EPR-type correlations in a bipartite system of
the photon and a spin-1/2 fermion of mass $m$. Such a system can
arise from a decay of another spin-1/2 fermion (of mass  $M$).

Let $q$ denote the four-momentum of the initial particle while $k$ and
$p$---the four-momenta of the final particles, i.~e.\ photon and
spin-$1/2$ fermion, respectively ($q=k+p$).

Let $\mathcal{H}_f$ be the carrier space of the irreducible
representation of the Poincar\'e group for fermion and
$\mathcal{H}_{\gamma}$ for photon. $\mathcal{H}_f$ is spanned by the
four-momentum operator eigenvectors, $\ket{p,\sigma}$, and
$\mathcal{H}_{\gamma}$ by $\ket{k,\lambda}$; $\sigma$ denotes here the
spin projection on $z$ axis and $\lambda$---helicity for the photon.
The above states do not transform in a manifestly covariant way [see
(\ref{eq:dzialanie_U(L)_na_baze}) and
(\ref{eq:dzialanie_U(L)_na_baze_foton}) in
Appendix~\ref{app:poincare}], but we can introduce states which
fulfill this requirement.

In the present paper we use the notation from our previous
papers \cite{CR2006, Caban2007_photons}. Summary of the notation
is given in Appendix~\ref{app:poincare}. 
To keep the notation as simple as possible we denote the
covariant states by kets with entries ordered as follows: a Lorentz
representation index ($\alpha$ -- bispinor, $\mu$ -- four-vector)
followed by a four-momentum, i.e.\ $\ket{\alpha,p}$,
$\ket{(\mu,\nu),k}$. Notice that the covariant vectors and the basis
vectors of the representation space of the Poincar\'e group
$\ket{p,\sigma}$, $\ket{k,\lambda}$, are distinguished by the order of
entries; explicit relationship between both kinds of vectors is given
below. 
\begin{itemize}
  \item For the spin-$1/2$  fermion
  \begin{subequations}
  \begin{gather}
    \label{eq:baza1_1/2}
    \ket{\alpha,p}=v_{\alpha\sigma}(p)\ket{p,\sigma},\\
    \label{eq:dzialanie_U(l)_na_baze2_1/2}
    U(\Lambda)\ket{\alpha,p}=
    \mathrm{D}(\Lambda^{-1})_{\alpha\beta}\ket{\beta,\Lambda p},
  \end{gather}
  \end{subequations}
where $\mathrm{D}(\Lambda)$ is the bispinor 
$D^{(1/2,0)}\oplus D^{(0,1/2)}$ representation of the Lorentz group
while $\Lambda$ denotes matrix of an arbitrary proper ortochronous
Lorentz transformation.
  \item For the photon
  \begin{subequations}
    \label{eq:kowariantne fotony}
     \begin{gather}
      \label{seq:baza_2_foton}
      \ket{(\mu,\nu),k}=f^{\mu\nu}_{\quad \lambda}(k)\ket{k,\lambda},\\
      \label{seq:transf. kowariantna dla fotonow}
      U(\Lambda)\ket{(\mu,\nu),k}=
      \Lambda^{-1\mu}_{\,\quad\mu'}\Lambda^{-1\nu}_{\,\quad\nu'}
      \ket{(\mu',\nu'),\Lambda k},
    \end{gather}
  \end{subequations}
where
 \begin{equation}
 \label{eq:f}
 f^{\mu\nu}_{\quad\lambda}(k)=
 k^{\mu}e^{\nu}_{\,\,\lambda}(k)-k^{\nu}e^{\mu}_{\,\,\lambda}(k).
 \end{equation}
\end{itemize}
Amplitudes $v_{\alpha\sigma}(p)$ and $e^{\mu}_{\,\,\lambda}(k)$ 
fulfill the Weinberg conditions
\cite{cab_Weinberg1964,cab_Weinberg1996}; for their 
explicit forms see  (\ref{eq:v}), (\ref{seq:e1}).

The Hilbert space of the hybrid system is a tensor product space
$\mathcal{H}_f\otimes \mathcal{H}_{\gamma}$ spanned by states
$\ket{p,\sigma}\otimes\ket{k,\lambda}$.  Because of transformation
rules (\ref{eq:dzialanie_U(l)_na_baze2_1/2}) and (\ref{seq:transf.
  kowariantna dla fotonow}), the manifestly covariant basis states of
the product space transform according to the
$(D^{\left(1/2,0\right)}\oplus D^{\left(0,1/2\right)})\otimes
\left(D^{(1,0)}\oplus D^{(0,1)}\right) =D^{\left(1/2,0\right)}\oplus
D^{\left(0,1/2\right)}\oplus D^{\left(1,1/2\right)}\oplus
D^{\left(1/2,1\right)}\oplus D^{\left(3/2,0\right)}\oplus
D^{\left(0,3/2\right)}$ representation of the Lorentz group. On the
other hand, since the considered hybrid state $\ket{\alpha; p,k}$ arises
from the decay of a spin-1/2 particle, it should transform according to the
formula analogous to (\ref{eq:dzialanie_U(l)_na_baze2_1/2})
 \begin{equation}
 \label{eq:dzialanie_U(l)_na_hybrydowe}
 U(\Lambda)\ket{\alpha;p,k}=
 \mathrm{D}(\Lambda^{-1})_{\alpha\beta}\ket{\beta;\Lambda
 p,\Lambda k}.
 \end{equation}
In terms of the states (\ref{eq:baza1_1/2}) and
(\ref{seq:baza_2_foton}), the basis vectors in the subspace of the
hybrid states take the form:
 \begin{equation}
 \ket{\alpha;p,k}=
 \left[A_{\mu\nu}(p,k)\right]_{\alpha\beta}
 \ket{\beta,p}\otimes\ket{(\mu,\nu),k}.
 \end{equation}
These vectors should transform according to
Eq.~(\ref{eq:dzialanie_U(l)_na_hybrydowe}), which imposes constraints
on $A_{\mu\nu}(p,k)$. By solving these constraints we find the following
distinct bases:
\begin{subequations}\label{seq:kp1kp2}
 \begin{eqnarray}
 \label{eq:kp1}
 \ket{\alpha;p,k}_1\!\!&=&
 \!\!\frac{1}{4}\left[\gamma_{\mu},\gamma_{\nu}\right]_{\alpha\beta}
 \ket{\beta,p}\otimes\ket{(\mu,\nu),k},\\
 \label{eq:kp2}
 \ket{\alpha;p,k}_2\!\!&=&
 \!\!(p_{\mu}\gamma_{\nu}-p_{\nu}\gamma_{\mu})_{\alpha\beta}
 \ket{\beta,p}\otimes\ket{(\mu,\nu),k},
 \end{eqnarray}
\end{subequations}
and
\begin{subequations}\label{seq:kp1kp2'}
 \begin{eqnarray}
 \label{dq:kp1'}
 \ket{\alpha;p,k}'_1&=&\gamma^5_{\alpha\beta}\ket{\beta;p,k}_1,\\
 \label{eq:kp2'}
 \ket{\alpha;p,k}'_2&=&\gamma^5_{\alpha\beta}\ket{\beta;p,k}_2.
 \end{eqnarray}
\end{subequations}
The parity operator, $\mathcal{\hat{P}}$, acts on the vectors in the 
following way: 
    \label{seq:parzystosci kp12}
    \begin{equation}
    \mathcal{\hat{P}}\ket{\alpha;p,k}_{1/2}=
    \zeta^*\gamma^0_{\alpha\beta}\ket{\beta;p^{\pi},k^{\pi}}_{1/2},
    \end{equation}
or
    \begin{equation}
    \mathcal{\hat{P}}\ket{\alpha;p,k}'_{1/2}=
    -\zeta^*\gamma^0_{\alpha\beta}
    \ket{\beta;p^{\pi},k^{\pi}}'_{1/2},
    \end{equation}
where $\zeta$ is the inner parity of final state fermion. States
(\ref{seq:kp1kp2}) correspond to the case, when the 
inner parities of decaying and product spin-1/2 fermion are equal, and
states (\ref{seq:kp1kp2'}) correspond to the case when decaying and
product spin-1/2 fermions have opposite parities.  

Hybrid states of a definite parity are combinations of the states
($\ref{seq:kp1kp2}$) or ($\ref{seq:kp1kp2'}$). 

Taking into account that the initial state of the decaying particle
must fulfill the Dirac equation 
 \begin{equation}
 \label{eq:dirac equation}
 (\hat{P}\gamma+M)\ket{\Psi}=0,
 \end{equation}
where $\hat{P}\gamma=\hat{P}_{\mu}\gamma^{\mu}$, with $\hat{P}_{\mu}$
being the four-momentum operator and $\gamma^{\mu}$ the Dirac matrices
(see Appendix~\ref{app:poincare}), 
we can determine the appropriate coefficients. As a result, we obtain:
    \begin{multline}
    \label{eq:q}
    \ket{{\Psi}}=
 \overline{\Psi}_{\alpha}\ket{\alpha;p,k}\\
 \equiv \overline{\Psi}_{\alpha}[(m+M)\ket{\alpha;p,k}_1+\ket{\alpha;p,k}_2],
 \end{multline}
or
 \begin{multline}
 \label{eq:q'}
 \ket{{\Psi}}'=
 \overline{\Psi}_{\alpha}\ket{\alpha;p,k}'\\
 \equiv \overline{\Psi}_{\alpha}[(m-M)\ket{\alpha;p,k}'_1+
 \ket{\alpha;p,k}'_2],
 \end{multline}
where $\Psi=[\Psi_{\alpha}]$ is a bispinor and
$\overline{\Psi}=\Psi^{\dag}\gamma^0$. 

Now, according to \cite{CR2005}, we choose $\Psi$ in such a way that
the spin reduced density matrix $\Psi\overline{\Psi}$ describes the
decaying particle with the polarization vector $\boldsymbol{\xi}$
 \begin{equation}
 \label{eq:macierz gestosci}
  \Psi\overline{\Psi}=
  \rho=\frac{1}{8}\left(\id+\frac{q\gamma}{M}\right)
  \left(\id+2\gamma^5\frac{w(q)\gamma}{M}\right)
  \left(\id+\frac{q\gamma}{M}\right),
 \end{equation}
where the four-vector $w$
 \begin{equation}
 \label{eq:w}
  w^0=\frac{\bold{q}\cdot\boldsymbol{\xi}}{2}, \quad \bold{w}=
 \frac{1}{2}\left(M\boldsymbol{\xi}+
 \frac{\bold{q}(\bold{q}\cdot\boldsymbol{\xi})}{M+q^0}\right)
 \end{equation}
is the mean value of the Pauli-Lubanski four-vector
$\hat{W}^{\mu}=
\tfrac{1}{2}\epsilon^{\nu\gamma\delta\mu}\hat{P}_{\nu}\hat{J}_{\gamma\delta}$
in the state $\rho$; $\hat{P}_{\nu}$ and $\hat{J}_{\gamma\delta}$ denote
the generators of the Poincar\'e group.

\section{Observables}
\label{s:observables}

In order to calculate the correlation function it is necessary to
introduce the spin operator for a relativistic massive particles and
the polarization operator for the photon.

Regarding the relativistic spin operator, several possibilities have
been discussed in literature (see
e.g.~\cite{Czachor1997_1,CW2003,Terno2003,RS2002,CR2005, CR2006,
  LY2004, LD2003, SL2004, TU2003_1, TU2003_2, czachor2010, bogolubov75}).
We restrict our considerations to the relativistic spin operator
 \begin{equation}
 \label{eq:definicja_operatora_spinu}
 \hat{\bold{S}}=
 \frac{1}{m}\left(\hat{\bold{W}}-
 \hat{W}^0\frac{\bold{\hat{P}}}{\hat{P}^0+m}\right),
 \end{equation}
which acts on one-particle states as follows:
 \begin{equation}
 \label{eq:dzialanie spinu na baze}
 \hat{\bold{S}}\ket{p,\sigma}=
 \frac{\boldsymbol{\sigma}_{\sigma'\sigma}}{2}\ket{p,\sigma'},
 \end{equation}
where $\sigma_i$ are the Pauli matrices.

The appropriately normalized polarization observable is given by the
obvious formula 
 \begin{equation}
 \label{eq:helicity}
 \hat{\bold{S}}(\theta)= 
 \frac{1}{2k^0\delta^3(0)}(\ket{\boldsymbol{\epsilon}_{\theta},k}
 \bra{\boldsymbol{\epsilon}_{\theta},k}-
 \ket{\boldsymbol{\epsilon}_{\theta_{\bot}},k}
 \bra{\boldsymbol{\epsilon}_{\theta_{\bot}},k}),
 \end{equation}
where
 \begin{equation}\label{eq:spolaryzowany foton}
 \ket{\boldsymbol{\epsilon}_{\theta},k}=
 \frac{1}{\sqrt{2}}(e^{i\theta}\ket{k,+1}+e^{-i\theta}\ket{k,-1})
 \end{equation}
is the state of the linearly polarized photon with four-momentum
$k$ and $\ket{\boldsymbol{\epsilon}_{\theta_{\bot}},k}=
\ket{\boldsymbol{\epsilon}_{\theta+\pi/2},k}$.
The vector $\ket{\boldsymbol{\epsilon}_{\theta},k}$ describes the
photon polarized in the plane spanned by the vectors $\bold{k}$ and
$\boldsymbol{\epsilon}_{\theta}\perp\bold{k}$, where: 
\begin{subequations}\label{seq:etheta}
 \begin{eqnarray}
 \label{eq:etheta}
 \boldsymbol{\epsilon}_{\theta}&=&
 \frac{1}{\sqrt{2}}\sum_{\lambda}\bold{e}_{\lambda}(k)
 e^{-i\lambda\theta},\\
 \label{eq:etheta_perp}
 \boldsymbol{\epsilon}_{\theta_{\perp}}&=&
 \frac{-i}{\sqrt{2}}\sum_{\lambda}\lambda\bold{e}_{\lambda}(k)
 e^{-i\lambda\theta}.
 \end{eqnarray}
\end{subequations}
The operator $\hat{\bold{S}}(\theta)$ acts on one-particle states
as follows:
 \begin{equation}
 \label{eq:dzialanie helicity na baze}
 \hat{\bold{S}}(\theta)\ket{k,\lambda}=
 \sum_{\lambda'}\frac{1-\lambda\lambda'}{2}
 e^{i(\lambda'-\lambda)\theta}\ket{k,\lambda'}.
 \end{equation}

\section{Correlation function} 
\label{s:correlation function}

Let us consider two distant observers, Alice and Bob, in
the same inertial frame, sharing the state $\ket{\Psi}.$ Let Bob
measure polarization of the photon and Alice---spin projection of the
fermion on some arbitrary direction $\bold{a}$, where $|\bold{a}|=1$.
As Alice assigns values $\pm 1$ instead of $\pm$1/2 to the outcomes of
her measurement, her observable is $2\bold{a}\cdot\hat{\bold{S}}$. The
correlation function takes the form
 \begin{equation}\label{eq:f_korelacji_definicja}
 C_{\Psi}(\theta,\bold{a},k,p)=
 \frac{\bra{\Psi} 2\bold{a}\cdot\hat{\bold{S}}\otimes
 \hat{\bold{S}}(\theta)\ket{\Psi}}{\bra{\Psi}\Psi\rangle}.
 \end{equation}
Inserting (\ref{eq:q}) into the above formula one gets:
\begin{widetext}
 \begin{multline}
 \label{eq:wzor na f korelacji}
  C_{\Psi}(\theta,\boldsymbol{\xi},\bold{a},k,p)=
  \frac{m}{32M (kp)^2}\sum_{\lambda}e^{-2i\lambda\theta}
  \tr\left\{[-(m+M)(k\gamma)(e_{\lambda}\gamma)+2(kp)(e_{\lambda}\gamma)
  -2(e_{\lambda}p)(k\gamma)](\id+2\gamma^5\frac{w(q)\gamma}{M})\right.\\
  \left.  \times[(m+M)(k\gamma)(e_{\lambda}\gamma)+
  2(kp)(e_{\lambda}\gamma)
  -2(e_{\lambda}p)(k\gamma)]v(p)
 (\bold{a}\cdot\boldsymbol{\sigma})^{\mathrm{T}}\overline{v}(p)
 \frac{}{}\!\right\}.
 \end{multline}
\end{widetext}
Now, taking into account the representation of the Dirac gamma matrices
(\ref{eq:dirac matrices}) and Eq.~(\ref{eq:v}), we have 
 \begin{multline}
 \label{eq:macierz spinu}
 v(p)(\bold{a}\cdot\boldsymbol{\sigma})^{\mathrm{T}}\overline{v}(p)=
 \frac{1}{2m}\left\{-
 \left(m\bold{a}+\frac{\bold{a}\cdot\bold{p}}{m+p^{0}}\bold{p}\right)
 \boldsymbol{\gamma}\gamma^{5}\right.\\
 +(\bold{a}\cdot\bold{p})\gamma^0\gamma^5\!
 \left.\!- i[(\bold{a}\times\bold{p})
 \boldsymbol{\gamma}]\gamma^0\!\!+
 \!\!\left(p^{0}\bold{a}\!-\!\frac{\bold{a}\cdot\bold{p}}{m+p^0}
 \bold{p}\right)\!\boldsymbol{\gamma}\gamma^0\!\gamma^5\right\}.
 \end{multline}
Using the trace properties of the Dirac matrices we finally get
 \begin{multline}
 \label{eq:kor_alpha_beta}
 C(\theta,\boldsymbol{\xi},\bold{a},k,p)=
 \frac{(\boldsymbol{\alpha}\cdot\boldsymbol{\epsilon_\theta})
 (\boldsymbol{\beta}\cdot\boldsymbol{\epsilon_\theta})
 -(\boldsymbol{\alpha}\cdot\boldsymbol{\epsilon_{\theta_{\perp}}})
 (\boldsymbol{\beta}\cdot\boldsymbol{\epsilon_{\theta_{\perp}}})}%
{(kp)^{2} (M+k^{0}+p^{0})},
 \end{multline}
where the polarization vectors are given by Eqs.~(\ref{seq:etheta})
and we use the following notation
 \begin{multline}
 \label{eq:alpha}
 \boldsymbol{\alpha}:=(kp)(M+p^{0}+k^{0})\boldsymbol{\xi}\\
 +\left[(M+p^{0})(\bold{k}\cdot\boldsymbol{\xi})
 -k^{0}(\bold{p}\cdot\boldsymbol{\xi})\right]\bold{p},
 \end{multline}
 \begin{equation}
  \label{eq:beta}
 \boldsymbol{\beta}:=(kp)\bold{a}+\left[(\bold{a}\cdot\bold{k})-
 \frac{k^{0}(\bold{a}\cdot\bold{p})}{m+p^{0}}\right]\bold{p},
 \end{equation}
(note that $kp=(M^2-m^2)/2$). The formula (\ref{eq:kor_alpha_beta}) is
valid also in the case of mixed states (when $|\boldsymbol{\xi}|<1$
instead of $|\boldsymbol{\xi}|=1$).  
The correlation function computed for the state $\ket{\Psi}^\prime$
(\ref{eq:q'}) differs from (\ref{eq:kor_alpha_beta}) by the overall
sign.   

Now, let us consider the correlation function in the center-of-mass
frame (c.m. frame), i.~e.~$\bold{p}=-\bold{k}$. 
The components $k^0$ and $p^0$ read
 \begin{equation}
  \label{k0p0cmf}
  k^0=\frac{M^2-m^2}{2M},\quad p^0=\frac{M^2+m^2}{2M},
 \end{equation}
and the components of the four-vector $w$ (\ref{eq:w}) take the form:
 \begin{equation}
  \label{eq:w w cmf}
  w^0=0,\quad \bold{w}=\frac{M\boldsymbol{\xi}}{2}.
 \end{equation}
In such a frame, the correlation function reduces to
 \begin{equation}
 \label{eq:kor_cmf}
 C_{\mathrm{c.m.}}(\theta,\boldsymbol{\xi},\bold{a})=
 (\bold{a}\cdot\boldsymbol{\epsilon_{\theta}})
 (\boldsymbol{\xi}\cdot\boldsymbol{\epsilon_{\theta}})-
 (\bold{a}\cdot\boldsymbol{\epsilon_{\theta_{\bot}}})
 (\boldsymbol{\xi}\cdot\boldsymbol{\epsilon_{\theta_{\bot}}}),
 \end{equation}
where the polarization vectors are defined by
Eqs.~(\ref{seq:etheta}). As we can see the function
(\ref{eq:kor_cmf}) does not depend on the value of particle momenta. 

In further considerations we have adopted the following
parametrization of the vectors $\bold{a}$ and $\bold{p}$: 
 \begin{align}
  \label{eq:parametryzacja}
  \bold{a} = \begin{pmatrix}
             \cos{\varsigma} \\
             \sin{\varsigma}\sin{\varphi} \\
             \sin{\varsigma}\cos{\varphi} 
             \end{pmatrix},
  \quad\quad\bold{p} = \begin{pmatrix}
             |\bold{p}|\cos{\psi} \\
             |\bold{p}|\sin{\psi}\\
             0
             \end{pmatrix}.
 \end{align}

\begin{figure}
\centering
    \includegraphics[width=1\columnwidth]{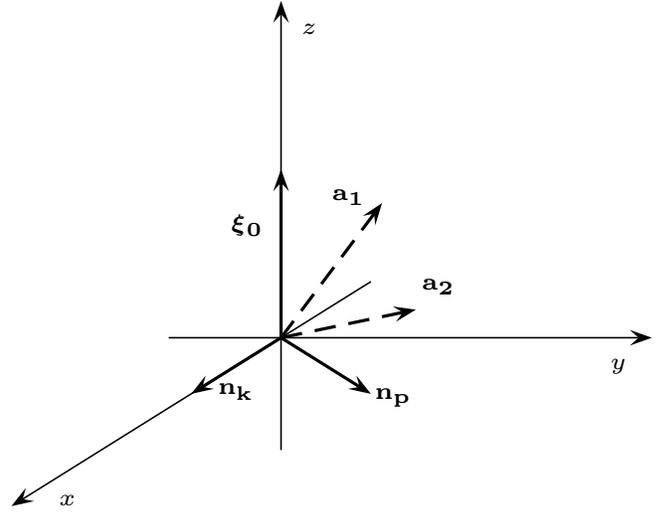}
    \caption{{Configuration in which the correlation function
        (\ref{eq:kor_ekstrema}) was calculated. Vector
        $\boldsymbol{\xi}$ [see Eq.~(\ref{eq:w})] lies on the $z$ axis,
        the photon propagates along the $x$ axis, the momentum of the
        spin-$1/2$ fermion ($\bold{n_p}$) lies on the $xy$ plane and
        the directions on which Bob measures spin projections of the
        fermion are denoted by $\bold{a_1}$ and $\bold{a_2}$. In the
        c.m. frame $\bold{n_k}=-\bold{n_p}$}.}
    \label{fig:konf}
\end{figure}

Without loss of generality we have assumed that the photon propagates along
$x$ axis, i.~e.\ $\bold{k}=(|\bold{k}|,0,0)$. In that case the
vectors (\ref{seq:etheta}) take the form 
 \begin{align}
  \label{eq:ethetaparametryzacja}
  \boldsymbol{\epsilon}_{\theta} = \begin{pmatrix}
                  0 \\
                 \sin{\theta} \\
                 \cos{\theta} 
                \end{pmatrix},
  \quad\quad\boldsymbol{\epsilon}_{\theta_{\perp}} =
                \begin{pmatrix}
                  0 \\
                 \cos{\theta}\\
                 - \sin{\theta}
                \end{pmatrix}.
 \end{align}
After setting $\boldsymbol{\xi}=(0,0,1)\equiv
\boldsymbol{\xi_0}$ (see Fig.~\ref{fig:konf}), 
the correlation function (\ref{eq:kor_alpha_beta}) takes the form
 \begin{multline}
  \label{eq:kor_ekstrema}
  C(\theta,\boldsymbol{\xi_0},\bold{a},k,p)=
  \sin\varsigma\cos(\varphi-2\theta)\\
  +\frac{\sqrt{x}}{\sqrt{x+1}-\sqrt{x}\cos\psi}
  \left[\cos\varsigma-
  \frac{\sqrt{x}}{\sqrt{x+1}+1}\left(\cos\varsigma\cos\psi\right.\right.\\
  \left.\left.+\sin\varsigma\sin\psi\sin\varphi\right)
  \frac{}{}\right]\sin\psi\sin 2\theta,
 \end{multline}
where $x=\left(\frac{|\bold{p}|}{m}\right)^2$.

\begin{figure}
\centering
    \includegraphics[width=1\columnwidth]{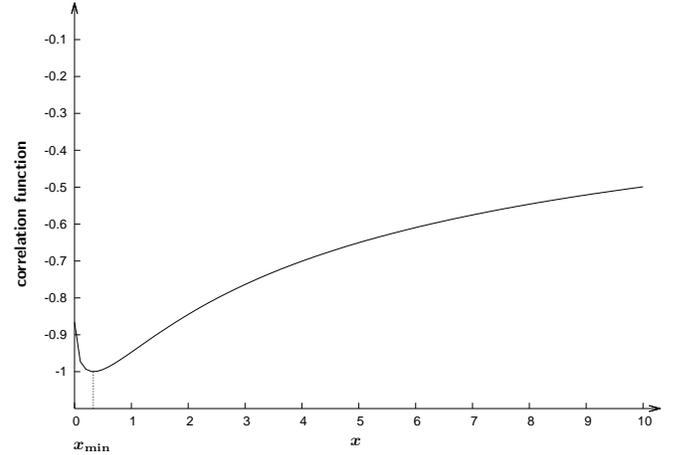}
    \caption{{Dependence of the correlation function
        $C(\theta,\boldsymbol{\xi_0},\bold{a},k,p)$ on
        $x=\left(\tfrac{|\bold{p}|}{m}\right)^2$ for
        $\varsigma=2\pi/3$, $\varphi=3\pi/2$,
        $\psi=\pi/3$ and $\theta=\pi/4$. The value of
        the minimum equals $-0.87$ ($x_{\mathrm{min}}=1/3).$}}
    \label{fig:minimum}
\end{figure}

\begin{figure}
\centering
    \includegraphics[width=1\columnwidth]{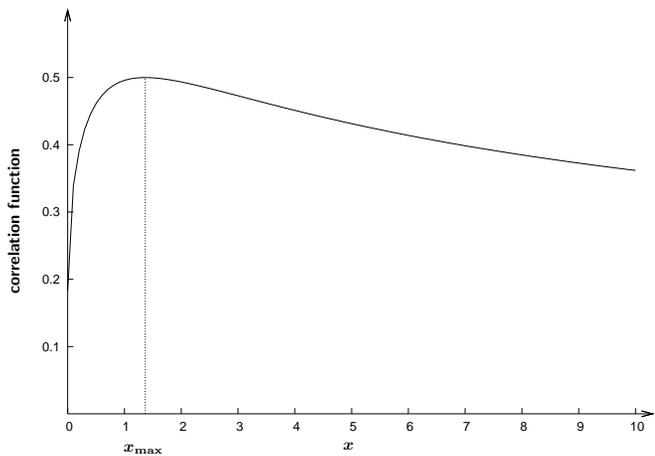}
    \caption{{Dependence of the correlation function
        $C(\theta,\boldsymbol{\xi_0},\bold{a},k,p)$ on
        $x=\left(\tfrac{|\bold{p}|}{m}\right)^2$ for
        $\varsigma=\varphi=\pi/4$ and
        $\psi=\theta=\pi/3$. The value at the maximum equals
        $1/2$ ($x_{\mathrm{max}}=1.36)$.}}
    \label{fig:maximum}
\end{figure}

We show the $x$-dependence of the function (\ref{eq:kor_ekstrema}) for
two sets of parameters:
$\varsigma=2\pi/3$, $\varphi=3\pi/2$,
$\psi=\pi/3$, $\theta=\pi/4$ (Fig \ref{fig:minimum})
and $\varsigma=\varphi=\pi/4$, $\psi=\theta=\pi/3$
(Fig \ref{fig:maximum}) The function (\ref{eq:kor_ekstrema}) has
extremum in both cases, a minimum and a maximum, respectively:
$C=-0.87$ for $x=1/3$ and $C=0.5$ for $x=1.36$. Such a property of the
correlation function in a bipartite systems of a vector bosons and
spin-1/2 fermions has already been reported by us \cite{ja2008,
Caban_2008_bosons_helicity, ja2009, CDKO_2010_fermion_helicity}.
Moreover, it also influences
the degree of violation of CHSH inequalities. One should notice,
that the velocities that enable observation of relativistic effects
are for a bipartite proton system of about $0.85 c$ \cite{smolinski}.
For hybrid system, extrema appear for $x=1/3$, i.~e.~for
$v=0.5c$, which is the border that was reached in RIKEN experiment
\cite{RIKEN}.

The ultra-relativistic limit of function (\ref{eq:kor_ekstrema}) reads
 \begin{multline}
  C_{ultrarel}(\theta,\boldsymbol{\xi_0},\bold{a})=
  \sin\varsigma\cos\varphi\cos 2\theta\\
  +\sin 2\theta(\cos \varsigma\sin\psi
  -\sin\varsigma\cos\psi\sin\varphi).
 \end{multline}
The non-relativistic limit  has the form
 \begin{equation}
  \label{eq:nierlelat}
  C_{nonrel}(\theta,\boldsymbol{\xi_0},\bold{a})=
  \sin\varsigma\cos(\varphi-2\theta),
 \end{equation}
which is exactly equal to the form of correlation function
(\ref{eq:kor_cmf}),
$\mathcal{C}_{\mathrm{c.m.}}(\theta,\boldsymbol{\xi},\bold{a})=
\mathcal{C}_{nonrel}(\theta,\boldsymbol{\xi_0},\bold{a})$, in
configuration we chose. 
Furthermore the function (\ref{eq:nierlelat}) has the same form as
(\ref{eq:nowe}) {up to normalization}.

\section{The CHSH inequality}
\label{s:bell}

In this section we consider the CHSH inequality and show, that it can be
violated in arbitrary reference frame. 

We search for the configuration, that maximally violates the CHSH
inequality which reads: 
 \begin{multline}
   \label{eq:chsh1}
   |C(\theta_1,\boldsymbol{\xi},\bold{a}_1,\bold{k},\bold{p})
   +C(\theta_1,\boldsymbol{\xi},\bold{a}_{2},\bold{k},\bold{p})\\
   +C(\theta_{2},\boldsymbol{\xi},\bold{a}_{2},\bold{k},\bold{p})
   -C(\theta_{2},\boldsymbol{\xi},\bold{a}_1,\bold{k},\bold{p})|\leq 2.
 \end{multline}
Just like in previous section, we assume that the photon propagates
along the $x$ axis and $\boldsymbol{\xi}$ is orientated along the $z$
axis (Fig.~\ref{fig:konf}). 

\begin{figure}
\centering
    \includegraphics[width=1\columnwidth]{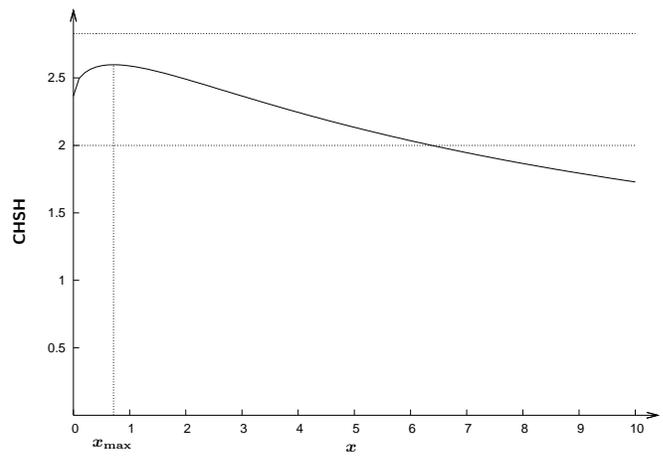}
    \caption{{Dependence of the left hand side of CHSH inequality
        (\ref{eq:chsh1}) on $x$ for $\psi=\pi/6$,
        $\varsigma_1=2\pi/3$, $\varphi_1=3\pi/2$,
        $\theta_1=2\pi/3$, $\varsigma_2=\pi/3$,
        $\varphi_2=\pi$ and $\theta_2=\pi/3$. It has maximum
        equal to $2.60$ at $x_{\mathrm{max}}=0.71.$ The inequality is
        violated for $x< 6.38$.}}
    \label{fig:chshmax}
\end{figure}
\begin{figure}
\centering
    \includegraphics[width=1\columnwidth]{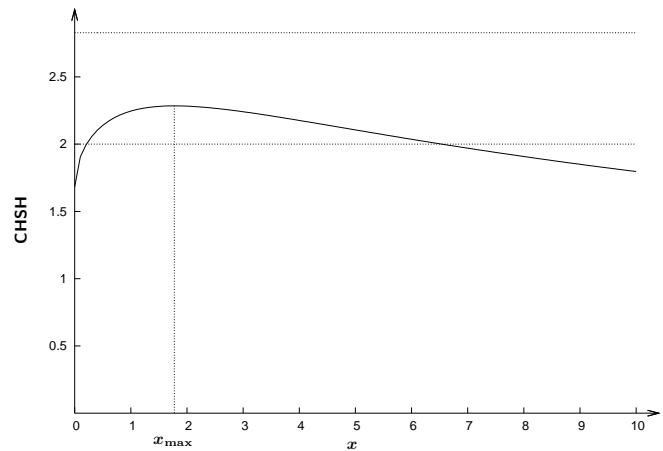}
    \caption{{Dependence of the left hand side of CHSH inequality
        (\ref{eq:chsh1}) on $x$ for $\psi=\varsigma_1={\pi}/{6}$,
        $\varphi_1={\pi}/{2}$, $\theta_1={3\pi}/{4}$,
        $\varsigma_2=\varphi_2={\pi}/{3}$ and
        $\theta_2={\pi}/{2}$. It has maximum equal to $2.28$ at
        $x_{\mathrm{max}}=1.77.$ The inequality is violated for
        $x\in(0.21,6.54)$.}}
    \label{fig:chshmax_1}
\end{figure}
\begin{figure}
\centering
    \includegraphics[width=1\columnwidth]{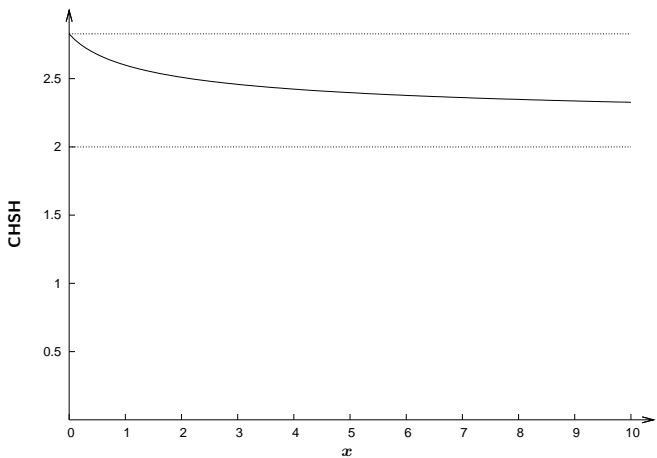}
    \caption{{Dependence of the left hand side of CHSH inequality
        (\ref{eq:chsh1}) on $x$ for $\psi={2\pi}/{3}$,
        $\varsigma_1={\pi}/{2}$, $\varphi_1={3\pi}/{4}$,
        $\theta_1={\pi}/{4}$, $\varsigma_2={\pi}/{2}$,
        $\varphi_2={\pi}/{4}$ and $\theta_2=0$. The inequality is
        violated in whole domain, the maximal violation occurs in
        non-relativistic case.}}
    \label{fig:chshmon}
\end{figure}

The correlation function (\ref{eq:kor_ekstrema}) violates the CHSH
inequalities, moreover, the degree of violation is related to the
existence of the extrema. This is shown in Fig.~\ref{fig:chshmax}
where the left hand side of (\ref{eq:chsh1}) is plotted versus the
variable $x$. We see, that the inequality is violated in
non-relativistic case, then the degree of its violation increases to
reach $2.60$ at $x_{\mathrm{max}}=0.71.$ Then it monotonically
decreases, and for $x>6.38$ the inequality is satisfied.  In
Fig.~\ref{fig:chshmax_1} the inequality is satisfied for the
non-relativistic case and violated for $x\in(0.21,6.54)$. It has
maximum equal to $2.28$ at $x_{\mathrm{max}}=1.77.$ In
Fig.~\ref{fig:chshmon}, the left hand side of the CHSH inequality is
always greater than $2$. In non-relativistic case the inequality is
maximally violated and then the degree of violation monotonically
decreases.

In c.m.\ frame, the CHSH inequality takes the form:
 \begin{multline}
  \label{eq:chsh2}
  2|\sin{\varsigma_1}\sin(\varphi_1-\theta_1-\theta_2)
  \sin(\theta_1-\theta_2)\\
  +\sin{\varsigma_2}\cos(\varphi_2-\theta_1-\theta_2)
  \cos(\theta_1-\theta_2)|\leq 2.
 \end{multline}
It follows from the above formula that in the relativistic case it is
possible to get the maximal violation of the CHSH inequality
($2\sqrt{2}$), for example when we set
$\varsigma_1=\varsigma_2=\pi/2$ and
$\theta_1-\theta_2=\pi/4$.  Thus, when
$\varphi_1-\theta_1-\theta_2=\pi/2$ and
$\varphi_2-\theta_1-\theta_2=0$ (e.g.~for $\varphi_1=3\pi/4$,
$\varphi_2=\pi/4$, $\theta_1=\pi/4$ and $\theta_2=0$)
the left hand side of (\ref{eq:chsh2}) is equal to $2\sqrt{2}$.

\section{Conclusions}
\label{s:conclusions}

We have constructed quantum state of a hybrid system (a massive
spin-1/2 fermion and the photon) arising from the decay of another
spin-1/2 fermion. The constructed state is characterized by the
polarization (Bloch) vector $\boldsymbol{\xi}$ [see
Eq.~(\ref{eq:macierz gestosci})]. 
We have calculated the correlation function in the
EPR-type experiment with hybrid system assuming that Alice measures
the spin projection of the fermion and Bob the polarization of the
photon. Next we have analyzed the correlation function in some
configurations and found that it can be a non-monotonic function of
the momentum of the fermion. Similar behavior of the correlation
function has been reported in the case of bipartite fermion (or boson)
system. However, in a hybrid system extrema occur for lower velocity
of the fermion ($0.5 c$) than in the two-fermion system ($0.85c$).
We have analyzed the CHSH inequality, too. We have found the
configuration in which the inequality is violated maximally. Moreover,
we have shown that the degree of violation of the CHSH inequality can
be a non-monotonic function of the fermion momentum. We have also
compared the results with non-relativistic case.

Let us note that a system fermion+photon can be produced also in the
Compton scattering. It seems that states prepared in this way are
easier to handle experimentally. Theoretical analysis of the
correlations in this case is more complicated and will be given in the
subsequent paper.

\begin{acknowledgments}
The authors are grateful to Jacek Ciborowski for interesting
discussion. 
This work has been supported by the University of Lodz and by the
Polish Ministry of Science and Higher Education under the contract N
N202 103738. 
P.~W. was supported by the European Union under the European Social
Fund: Human - best investment.
\end{acknowledgments}

\appendix

\section{Poincar\'e representations
 of spin-1/2 fermion and a
  photon} 
\label{app:poincare}

For readers convenience and to establish the notation, we recall some
basic facts and formulas necessary to follow the paper.

\subsection{Spin-1/2 fermion}

The notation and formalism we use in the case of spin-1/2 particles is
explained wider in our previous paper \cite{CR2006}. Here
we briefly recall the most important points.
For spin-1/2 fermion, the carrier space of irreducible representation
of the Poincar\'e group is spanned by the four-momentum operator
eigenvectors $\ket{p,\sigma}$, where $p^2=m^2$, with $m$ denoting the
mass of the particle, and $\sigma=\pm 1/2$.  Their transformation rule
reads
 \begin{equation}
 \label{eq:dzialanie_U(L)_na_baze}
 U(\Lambda)\ket{q,\sigma}=
 \mathcal{D}_{\sigma'\sigma}(R(\Lambda,q))\ket{\Lambda q,\sigma'},
 \end{equation}
where $\mathcal{D}$ is the spin $1/2$ fundamental
representation of SU(2), and the Wigner rotation
$R(\Lambda,q)\in$ SO(3) is defined as 
$R(\Lambda,q)=L_{\Lambda q}^{-1}\Lambda L_q$.
We use Lorentz-covariant normalization
 \begin{equation}
 \label{eq:normalizacja_wektorow_bazowych}
 \bracket{p,\sigma}{p',\sigma'}=
 2p^0\delta^3(\bold{p}-\bold{p}^{\prime})\delta_{\sigma\sigma'}.
 \end{equation}

Consistency of Eqs.~(\ref{eq:dzialanie_U(L)_na_baze},
\ref{eq:baza1_1/2}, \ref{eq:dzialanie_U(l)_na_baze2_1/2}) leads to
Weinberg condition 
 \begin{equation}
  \label{eq:warunek weinberga}
  \mathrm{D}(\Lambda)v(p)\mathcal{D}^{\mathrm{T}}(R(\Lambda,p))
  =v(\Lambda p).
 \end{equation}
For $\mathrm{D}$ being a bispinor representation
$D^{\left(1/2,0\right)}\oplus D^{\left(0,1/2\right)}$ of the Lorentz
group, we can find (see e.g.~\cite{CR2005})
 \begin{equation}
 \label{eq:v}
  v(p)=\frac{1}{2\sqrt{1+\tfrac{p^0}{m}}}
           \begin{pmatrix}
             (\id+\tfrac{1}{m}p^{\mu}\sigma_{\mu})\sigma_2 \\
             (\id+\tfrac{1}{m}(p^{\pi})^{\mu}\sigma_{\mu})\sigma_2 \\
           \end{pmatrix},
 \end{equation}
where $\sigma_{0}\equiv \id$, $\sigma_i$ are standard Pauli matrices
and $p^{\pi}=(p^0,-\bold{p})$.
The amplitudes (\ref{eq:v}) fulfill
\begin{subequations}
    \label{seq:warunki v}
    \begin{eqnarray}
    \label{eq:war1}
    v(p)\overline{v}(p)&=&\frac{1}{2m}(m+p\gamma),\\
    \label{eq:war2}
    \overline{v}(p)v(p)&=& \openone,
    \end{eqnarray}
\end{subequations}
where $p\gamma=p_{\mu}\gamma^{\mu}$ and
$\overline{v}(p)=v^{\dag}(p)\gamma^0$ stands for Dirac conjugation. We
use the following explicit representation of gamma matrices 
 \begin{equation}
 \label{eq:dirac matrices}
    \gamma^0=\begin{pmatrix}
               0 & \id \\
               \id & 0
               \end{pmatrix},\quad
    \boldsymbol{\gamma}=\begin{pmatrix}
               0 & -\boldsymbol{\sigma} \\
               \boldsymbol{\sigma}& 0
               \end{pmatrix},\quad
    \gamma^5=\begin{pmatrix}
               \id & 0 \\
               0 & -\id
               \end{pmatrix},
 \end{equation}
{where $\boldsymbol{\sigma}=(\sigma_1,\sigma_2,\sigma_3)$.
It holds
 \begin{equation}
 \label{eq:norma kret p}
 \bra{\delta,p}\gamma^{0}_{\delta\alpha}\ket{\beta,p'}=
 \frac{p^0}{m}\delta^3(\bold{p}-\bold{p}^{\prime})
 \left[m+p\gamma\right]_{\beta\alpha}.
 \end{equation}

The action of the space inversion operator $\hat{\mathcal{P}}$ on the
basis states is given by 
 \begin{equation}
 \label{eq:parzystosc_fermion_noncov}
  \mathcal{\hat{P}}\ket{p,\sigma}=\zeta^*\ket{p^{\pi},\sigma},
 \end{equation}
consequently for covariant states defined by (\ref{eq:baza1_1/2})
 \begin{equation}
 \label{eq:parzystosc_fermion_cov}
 \mathcal{\hat{P}}\ket{\alpha,p}=
 \zeta^*\gamma^0_{\alpha\beta}\ket{\beta,p^{\pi}},
 \end{equation}
with $\zeta$ denoting the inner parity of the particle.

\subsection{Photon}
\label{app:photon}

The notation and formalism we use in the case of photons is
explained wider in our previous paper \cite{Caban2007_photons}. Here
we briefly recall the most important points.
The carrier space of the irreducible
{photon} representation of the Poincar\'e group is spanned
by four-momentum eigenstates $\ket{k,\lambda}$ with $\lambda=\pm1$
denoting helicity and $k^2=0$.
Their transformation rule reads
 \begin{equation}
 \label{eq:dzialanie_U(L)_na_baze_foton}
 U(\Lambda)\ket{k,\lambda}=
 e^{i\lambda\psi(\Lambda,k)}\ket{\Lambda k,\lambda},
 \end{equation}
where $e^{i\lambda\psi(\Lambda,k)}=U(R(\Lambda,k))$.

They are normalized as follows:
 \begin{equation}
 \label{eq:normalizacja_wektorow_bazowych_foton}
 \bracket{k,\lambda}{k',\lambda'}=
 2p^0\delta^3(\bold{k}-\bold{k}^{\prime})\delta_{\lambda\lambda'}.
 \end{equation}
The vectors $\ket{k,\lambda}$ can be generated from standard vector
$|\tilde{k},\lambda\rangle$, with $\tilde{k}=(1,0,0,1)$. We have
$\ket{k,\lambda}=U(L_k)|\tilde{k},\lambda\rangle$, where
$L_k=R_{\bold{n}_\bold{k}}B(k^0)$. $B(k^0)$ is a pure Lorentz boost
taking vector $\tilde{k}$ into $k^0\tilde{k}$ and $R_{{\bold{n_k}}}$
is rotation transforming vector $\tilde{k}$ into
$(1,\bold{n}_{\bold{k}})$, where
$\bold{n}_{\bold{k}}=\bold{k}/|\bold{k}|$.

The explicit form of amplitudes $e_{\lambda}(k)$, which define the
covariant state (\ref{eq:kowariantne fotony}) reads 
    \begin{equation}
    \label{seq:e1}
    [e_{\mu\lambda}(k)]=\frac{1}{\sqrt{2}} R_{\bold{n}_{\bold{k}}}
                                                  \begin{pmatrix}
                                                    0 \\
                                                    -1 \\
                                                    i\lambda \\
                                                    0 \\
                                                  \end{pmatrix}.
    \end{equation}
The general form of $R_{\bold{n}_{\bold{k}}}$ is \cite{ja2008}
 \begin{equation}
 \label{eq:RNK}
  R_{\bold{n}_{\bold{k}}}=\left(\begin{array}{c|c}
    1 & \boldsymbol{0}^{\rm{T}} \\\hline
    \boldsymbol{0} & \bold{a}_{\bold{k}}|\bold{n}_{\bold{k}}
    \times \bold{a}_{\bold{k}}| \bold{n}_{\bold{k}}
    \end{array}\right),
 \end{equation}
where $\bold{a}_{\bold{k}}\bot\bold{n}_{\bold{k}}$,
$|\bold{a}_{\bold{k}}|=1$.

The choice of $\bold{a}_{\bold{k}}$ is the matter of convention.
Without loss of generality we can assume that photon propagates along
$x$ axis and then choose $\bold{a}_{\bold{k}}=(0,0,1)$, which is the
standard choice we use in this paper.

The action of the space inversion operator $\hat{\mathcal{P}}$ on the
basis states is: 
 \begin{equation}
 \label{eq:parzystosc_foton_noncov}
 \mathcal{\hat{P}}\ket{k,\lambda}=
 \chi_{\lambda}(k)\ket{k^{\pi},-\lambda},
 \end{equation}
and consequently for covariant states defined by (\ref{seq:baza_2_foton})
 \begin{equation}
 \label{eq:parzystosc_foton_cov}
 \mathcal{\hat{P}}\ket{(\mu,\nu),k}=
 \eta^{\mu\mu}\eta^{\nu\nu}\ket{(\mu,\nu),k^{\pi}},
 \end{equation}
with
 \begin{equation}
 \bold{e}_{\lambda}(k)=
 -\chi_{-\lambda}(k^{\pi})\bold{e}_{-\lambda}(k^{\pi})
 \end{equation}
and
 \begin{equation}
 \chi_{\lambda}(k)\chi_{-\lambda}(k^{\pi})=1.
 \end{equation}
In the convention used above $\chi_{\lambda}(k)=-1$.

Coefficients $f^{\mu\nu}_{\lambda}$ defined by Eq.~(\ref{eq:f}) fulfill
 \begin{multline}
  \label{eq:relacja}
  f^{*\mu\nu}_{\lambda}f^{\mu'\nu'}_{\lambda} = 
  \eta^{\mu\nu'}k^{\nu}k^{\mu'} + 
  \eta^{\nu\mu'}k^{\mu}k^{\nu'} \\
  - \eta^{\mu\mu'}k^{\nu}k^{\nu'} - 
  \eta^{\nu\nu'}k^{\mu}k^{\mu'},
 \end{multline}
where $\eta=\mathrm{diag}(1,-1,-1,-1).$

\section{Non-relativistic case}
\label{app:nonrel} 

As a non-relativistic analog of a hybrid system we take the system
consisting of a spin-1/2 fermion and a spin-1 massive boson. The total
spin of the system is equal to 1/2. Using the Clebsh-Gordan
coefficients we can write down two linearly independent states of the
system with total spin equal to 1/2
 \begin{subequations}
 \begin{eqnarray}
 \label{eq:nierelatywistycznyy}
    \ket{\uparrow}&=&
    \sqrt{\frac{2}{3}}\ket{-1/2}_f\otimes\ket{1}_b
    -\sqrt{\frac{1}{3}}\ket{1/2}_f\otimes\ket{0}_b,\notag\\&&\\
    \ket{\downarrow}&=&\sqrt{\frac{1}{3}}\ket{-1/2}_f\otimes\ket{0}_b
    -\sqrt{\frac{2}{3}}\ket{1/2}_f\otimes\ket{-1}_b,\notag\\&&
 \end{eqnarray}
 \end{subequations}
where $\ket{\sigma}_{f/b}$ stands for the state of a fermion/boson with
spin projection on $z$ axis equal to $\sigma$. 

Therefore, the general state of the system with total spin equal to
1/2 has the following form
 \begin{equation}
 \label{eq:nierelatywistyczny}
 \ket{\psi}=\alpha\ket{\uparrow}+\beta\ket{\downarrow}.
 \end{equation}
The density matrix corresponding to the above state reads
 \begin{multline}
 \label{eq:matrix_psi}
 \ket{\psi}\bra{\psi} = \frac{1}{2}\big[ (\ket{\uparrow}\bra{\uparrow}+
 \ket{\downarrow}\bra{\downarrow}) +
 \xi_1(\ket{\downarrow}\bra{\uparrow}
 +\ket{\uparrow}\bra{\downarrow}) \\
 +i\xi_2(\ket{\downarrow}\bra{\uparrow}
 -\ket{\uparrow}\bra{\downarrow}) +
 \xi_3(\ket{\uparrow}\bra{\uparrow}-
 \ket{\downarrow}\bra{\downarrow})\big],
 \end{multline}
where the components of the Bloch vector
$\boldsymbol{\xi}=(\xi_1,\xi_2,\xi_3)$ are connected with the
coefficients $\alpha,\beta$ from Eq.~(\ref{eq:nierelatywistyczny})
by
 \begin{subequations}
 \begin{eqnarray}
    |\alpha|^2&=&(1+\xi_3)/2,\\
    |\beta|^2&=&(1-\xi_3)/2,\\
    \alpha\beta^*&=&(\xi_1-i \xi_2)/2.
 \end{eqnarray}
 \end{subequations}
Note, that in the subspace spanned by the vectors $\ket{\uparrow}$ and
$\ket{\downarrow}$, matrix elements of the operators 
$(\ket{\uparrow}\bra{\uparrow}+
 \ket{\downarrow}\bra{\downarrow})$, 
$(\ket{\downarrow}\bra{\uparrow}
 +\ket{\uparrow}\bra{\downarrow})$,
$i(\ket{\downarrow}\bra{\uparrow}
 -\ket{\uparrow}\bra{\downarrow})$,
$(\ket{\uparrow}\bra{\uparrow}-
 \ket{\downarrow}\bra{\downarrow})$
are the same as the matrix elements of the Pauli matrices $\sigma_0$,
$\sigma_1$, $\sigma_2$, and $\sigma_3$.

We want to calculate the non-relativistic correlation function 
 \begin{equation}
  \label{eq:nonrel_f_kor}
 \frac{\bra{\psi}2\bold{a}\cdot\boldsymbol{\hat{S}}
 \otimes\hat{S}_\theta\ket{\psi}}{\bracket{\psi}{\psi}},
 \end{equation}
where $\bold{a}\cdot\boldsymbol{\hat{S}}$ is an observable
measuring spin projection of a spin-1/2 particle on the direction
$\bold{a}$, i.~e.
 \begin{equation}
 \bold{a} \cdot\boldsymbol{\hat{S}} \ket{\sigma}_f = \frac{1}{2} 
 \bold{a} \cdot\boldsymbol{\sigma}_{\lambda\sigma} \ket{\lambda}_f.
 \end{equation}
$\hat{S}_\theta$ is the polarization observable. It is
convenient to define the polarization observable in terms of
helicity basis $\ket{\bold{n_k},\lambda}$, with
$\bold{n_k}=\bold{k}/|\bold{k}|$ denoting the direction of
spin-1
particle momentum and $\lambda$ standing for its helicity. The
helicity basis can be expressed by means of spin basis as follows:
 \begin{equation}
 \label{eqn:helicity_spin}
 \ket{\bold{n_k},\lambda} = 
 \mathcal{D}^{(1)}_{\sigma\lambda}(R_{\bold{n_k}}) \ket{\sigma}_b,
 \end{equation}
where $R_{\bold{n_k}}$ is given by Eq.~(\ref{eq:RNK}) and 
$\mathcal{D}^{(1)}$ denotes standard spin-1 representation of the
rotation group (see e.g. \cite{Caban_2008_bosons_helicity}).
In the above basis, the polarization observable is defined in analogy
to Eq.~(\ref{eq:helicity}) by
 \begin{equation}
 \hat{S}_\theta=\ket{\boldsymbol{\epsilon}_{\theta},\bold{n_k}}\!
 \bra{\boldsymbol{\epsilon}_{\theta},\bold{n_k}}-
 \ket{\boldsymbol{\epsilon}_{\theta_{\bot}},\bold{n_k}}\!
 \bra{\boldsymbol{\epsilon}_{\theta_{\bot}},\bold{n_k}},
 \end{equation}
where
 \begin{equation}
 \ket{\boldsymbol{\epsilon}_{\theta},\bold{n_k}}=
 \frac{1}{\sqrt{2}}(e^{i\theta}\ket{\bold{n_k},+1}+
  e^{-i\theta}\ket{\bold{n_k},-1})
 \end{equation}
and $\theta_{\bot}=\theta+\pi/2$.

{Let us now consider a configuration, when the three-momentum of
the boson is along the $x$ axis [$\bold{n_k}=(1,0,0)$], and
$\bold{a_k}=(0,0,1)$, where $\bold{a_k}$ is the vector defining
rotation (\ref{eq:RNK}). In such a configuration
\begin{subequations}
 \begin{align}
 \ket{\bold{n_k},1} & = 
 -\frac{1}{2} (\ket{1}_b+\sqrt{2}\ket{0}_b+\ket{-1}_b),\\
 \ket{\bold{n_k},0} & = \frac{1}{\sqrt{2}} (\ket{1}_b-\ket{-1}_b),\\
 \ket{\bold{n_k},-1} & =
 \frac{1}{2} (\ket{1}_b-\sqrt{2}\ket{0}_b+\ket{-1}_b).
 \end{align}
\end{subequations}
Therefore
 \begin{multline}
 \ket{\boldsymbol{\epsilon}_{\theta},\bold{n_k}} =
  -\frac{i\sin\theta}{\sqrt{2}}
 (\ket{1}_b)+\ket{-1}_b) - \cos\theta \ket{0}_b.
 \end{multline}

In this case the correlation function in the state
(\ref{eq:nierelatywistyczny}) reads}
 \begin{multline}
 \label{eq:kor_nierel}
 C(\theta,\boldsymbol{\xi},\bold{a})=\\
 \frac{2}{3}\left[(a_3\xi_3-a_2\xi_2)
 \cos 2\theta+(a_2\xi_3+a_3\xi_2)\sin 2\theta\right].
 \end{multline}
When we use parametrization (\ref{eq:parametryzacja}) it reduces to
 \begin{equation} 
 \label{eq:nowe}
 C(\theta,\boldsymbol{\xi_0},\bold{a})=
 \frac{2}{3}\sin\varsigma\cos(\varphi-2\theta),
 \end{equation}
Note that the factor $2/3$ in formula (\ref{eq:kor_nierel}) appears
because the probabilities of that Bobs measurement outcome equals $0$,
enters the correlation function with measure $0$. If we normalized all
probabilities entering the correlation function to $1$, the factor
would be $1$, just as in the relativistic case, where the probability
of Bob getting the outcome $0$ vanishes.

\bibliographystyle{apsrev}


\end{document}